\documentclass[11pt,twoside]{article}
\usepackage{asp2006}
\usepackage{epsfig}
\usepackage{epsf}
\usepackage{graphics}
\usepackage{lscape}
\def\edcomment#1{\iffalse\marginpar{\raggedright\sl#1\/}\else\relax\fi}
\reversemarginpar

\begin{document}
\title{X- and Gamma-Ray Emission from the Galactic Center }
\author{V.A. Dogiel$^1$,  K.-S. Cheng$^2$, D.O. Chernyshov$^{1}$, W.-H. Ip$^3$, C.-M. Ko$^3$  and , V. Tatischeff$^4$} \affil{$^1$P.N.Lebedev Institute, Leninskii pr, 53, 119991
Moscow, Russia}\affil{$^2$University of Hong Kong, Hong Kong,
China}
    \affil{$^3$Institute of Astronomy, National Central University,
Chung-Li, Taiwan} \affil{$^4$Centre de Spectrom\'etrie
Nucl\'eaire, France}

\begin{abstract}
We discuss the origin of continuum and line X-ray emission
observed in the direction the Galactic Center. We predict a
significant flux of de-excitation gamma-ray lines in this
direction, which can be produced by subrelativistic protons
generated by accretion processes.
\end{abstract}

\vspace{-0.5cm}
\section{Introduction}
Galactic center (GC) is a harbour of high energy activity which is
observed in different ranges of electromagnetic waves.

 The inner 10 pc of our Galaxy is a source  of the very high energy
 $\gamma$-rays ($E_\gamma>100$ GeV)   coincident with the position of
 the supermassive black hole Sgr A$^\ast$ \citep{acero}.

The EGRET telescope \citep[see][]{mayer} found a gamma-ray flux
toward the Galactic Center of the order of $2\cdot 10^{37}$ erg
s$^{-1}$ for energies E$ > 500$ MeV in an error circle of 0.2
degree radius.

The INTEGRAL team measured a 511 keV line of width 2.2 keV, with a
flux $1.01\cdot 10^{-3}$ photons cm$^{-2}$ s$^{-1}$. The bulge
annihilation emission is highly symmetric with an extension of 5-8
degree. \citep[see e.g.,][]{chur_INTEGRAL}.

Last Suzaku observations of \citet{koya2} found a clear evidence
for a hot plasma in the GC with the diameter about 20  acrminutes
(i.e. $\sim 50–-60$ pc). The total X-ray flux from this region in
the range 2 to 10 keV is $F_X\sim 2\cdot 10^{36}$ erg s$^{-1}$,
and the total energy of plasma in this region is about $3\cdot
10^{52}$ erg. Such a high plasma temperature is surprising, since
it could not be gravitationally confined and a very high amount of
energy ($\sim 10^{42}$ erg s$^{-1}$ is required to maintain the
plasma outflow \citep[see, e.g.][]{koya1}). This energy supply
cannot be produced by SN explosions and other more powerful
sources of energy are required to support the energy balance
there.

Intensive emission in X-ray iron lines is observed from the
Galactic center, which is often explained that the gas there was
exposed in the past by sources of intensive X-ray emission e.g.,
from a supernova or from the galactic nucleus \citep{sun1,koya1}.

 The IBIS/ISGRI imager on the INTEGRAL observatory detected for the first time a hard continuum X-ray emission
  located within $1^\prime$ of Sgr A* over the energy range 20-100 keV \citep{belanger}.

 Latter \citet{koya2} also  found that
the continuum flux from the GC contained an additional hard
component.  \citet{yuasa} performed analysis of Suzaku data and
showed a prominent hard X-ray emission in the range from 14 to 40
keV whose spectrum is a power law with the spectral index ranging
from 1.8 to 2.5. The total luminosity of the power-law component
from the central region  is about $4\cdot 10^{36}$ erg s$^{-1}$.
The spatial distribution of hard X-rays correlates with the
distribution of hot plasma.

We assume that this activity of the Galactic center in different
ranges of waves has common origin, namely, it is due to processes
of star accretion onto the central black hole.

In \citet{cheng1,cheng2} we discussed  the origin of the 511 keV
annihilation flux from the GC region and production of continuum
gamma-ray emission in the range $E_\gamma>100$ MeV. The origin of
the 511 keV line emission from the GC region is supposed to be due
to annihilation of secondary positrons generated by $p-p$
collisions. Below we present a model of X-ray and de-excitation
gamma-ray line emission which can also be produced by this
activity.
\section{Energy Release}

Energy release of black holes due to processes of accretion can be
observed in different wave ranges.

{\it X-rays.} The maximum  flux of X-rays can be estimated from
the equation for the Eddington emissivity,
\begin{equation}
L_{Edd}\simeq \frac{4\pi M_{bh}Gm_pc}{\sigma_T}
\end{equation}
For a black hole with the mass $M_{bh}\sim 10^6M_\odot$ it give
about $L_{Edd}\sim 10^{44}$erg s$^{-1}$.

{\it Flux of relativistic charged particles.} Flux of relativistic
particles in the form of jets (electrons or protons). The origin
of relativistic protons in jets is still rather speculative but we
have evidences in favour of their production near black holes
\citep[see][]{auger,ist}. The energy carried away by relativistic
protons is estimated as \citep[see][]{cheng1}
\begin{eqnarray}
\Delta{E_p} \sim 5 \times 10^{51}(\eta_p/10^{-2})
(M_{\ast}/M_{\odot}) \mbox{erg}. \label{erg}\,\,
\end{eqnarray}
where $\eta_p$ is the conversion efficiency  from accretion power
 into the the energy of jet motion and $M_\ast$ is the star mass.

 {\it Flux of subrelativistic protons.} Once passing the pericenter, the star is tidally disrupted into a
very long and dilute gas stream. The outcome of tidal disruption
is that some energy is extracted out of the orbit to unbind the
star and accelerate the debris. Initially about 50\% of the
stellar mass becomes tightly bound to the black hole , while the
remainder 50\% of the stellar mass is forcefully ejected
\citep[see, e.g.][]{ayal}. The kinetic energy carried by the
ejected debris is a function of the penetration parameter $b^{-1}$
and can significantly exceed that released by a normal supernova
($\sim 10^{51}$~erg) if the orbit is highly penetrating
\citep[see][]{alex05},
\begin{equation}\label{energy}
  W\sim 4\times 10^{52}\left(\frac{M_\ast}{M_\odot}\right)^2
  \left(\frac{R_\ast}{R_\odot}\right)^{-1}\left(\frac{M_{\rm bh}/M_\ast}{10^6}\right)^{1/3}
  \left(\frac{b}{0.1}\right)^{-2}~\mbox{erg}\,.
\end{equation}
For the star capture time $\tau_s\ga 10^{4}$ years
\citep[see][]{alex05}
 it gives a power input $W \la 3\cdot 10^{42}$ erg s$^{-1}$.
 The mean
kinetic energy per escaping nucleon is given by
\begin{equation}\label{esc}
  E_{\rm esc}\sim 42 \left(\frac{\eta}{0.5}\right)^{-1} \left(\frac{M_\ast}{M_\odot}\right)
  \left(\frac{R_\ast}{R_\odot}\right)^{-1}\left(\frac{M_{\rm bh}/M_\ast}{10^6}\right)^{1/3}
  \left(\frac{b}{0.1}\right)^{-2}~\mbox{MeV}\,,
\end{equation}
where $\eta M_\ast$ is the mass of escaping material, $b$ is the
ratio of $r_p$ - the periapse distance (distance of closest
approach) to the tidal radius $R_T$. For the black-hole mass
$M_{\rm bh}=4.31 \times 10^6~M_{\odot}$ the energy of escaping
particles is $E_{\rm esc} \sim 68 (\eta /0.5)^{-1}
(b/0.1)^{-2}~\mbox{MeV nucleon$^{-1}$}$ when a one-solar mass star
is captured.

The dissipation times of these energy components released in
accretion processes are quite different: if the duration time of
X-ray emission $\tau_X$ from a black hole is about hundred years,
the characteristic lifetime of relativistic protons can be (at
some conditions) as small as $\tau_{rp}< 10^4$ years while that of
100 MeV subrelativistic protons is of the order of $\tau_{srp}\sim
10^7$ years \citep[see][]{cheng1,dog_pasj}, i.e.
$\tau_X,\tau_{rp}<\tau_s\ll\tau_{srp}$. In this respect processes
concerned with subrelativistic protons  can be considered as
quasi-stationary  (unlike that of X-rays or relativistic protons).
Emission generated by subrelativistic protons is presented below.
\section{Origin of X-ray Emission from the Galactic Center and
Production of De-Excitation Gamma-Ray Lines There}

{\it Thermal X-ray continuum}. The average power of energy release
from accretion in the form of subrelativistic protons is about
$10^{42}$ erg s$^{-1}$ and the average energy of these protons is
about 100~MeV. These  protons transform their energy into plasma
heating by ionization losses. As derived by \citet{{sun1},koya1}
just this energy release is necessary to heat the plasma  up to
temperatures about $6-10$~keV, just as observed
 \citep[see][]{dog_pasj}.

{\it Non-thermal X-ray continuum}.The inverse bremsstrahlung
losses of the protons  produce  a non-thermal X-ray flux in the
range above 10 keV. For the parameters of accretion the inverse
bremsstrahlung flux of protons is about $3\times 10^{36}$ erg
s$^{-1}$, \citep[see][]{dog_pasj} i.e. about the flux observed by
Suzaku from the GC in the $14-40$~keV band \citep{yuasa}.

{\it Flux of 6.4 keV iron line from molecular clouds}. As we
mentioned above the GC is active not only in X-ray continuum but
 high fluxes of 6.4 keV iron line are observed in the direction of  molecular
clouds in the GC. One of the interpretation is that these flux
arise due to the K-absorption of photons with energies E $> 7.1$
keV by dense molecular clouds  irradiated by a nearby X-ray
source, which was active in the recent past ($\sim 300–-400$ years
ago) but is almost unseen at present \citep{{sun1},koya1}. In this
case the 6.4 keV emission from the Galactic center is observed by
chance and its duration cannot be longer than several ten years.

An alternative interpretation of the origin of the K² fluorescent
line in the Galaxy is its excitation by subrelativistic charged
particles, i.e., by electrons \citep[see e.g.][]{yus1} or protons
\citep[see][]{dog4}.

The proton distribution inside the cloud depends on the processes
of proton penetration into dense neutral gas. As it was shown in
\citet{dog87}  strong fluctuations of the magnetic field are
induced by the observed gas turbulence inside molecular clouds
that makes cosmic ray propagation there similar to diffusion with
a relatively small diffusion coefficient. Therefore,
subrelativistic protons are able to fill a part of  of the cloud
volume, i.e. a surface envelope with the thickness $\la 1$ pc. Our
calculations show the rate of ionization in  clouds by
subrelativistic protons is strongly nonuniform. It is very high at
the cloud surface, $\zeta\ga 10^{-13}$ s$^{-1}$H$^{-1}$, but
decreases almost to zero with the distances  from the surfaces.

 For the parameters of quasi-stationary flux of
subrelativistic protons produced by accretion one can estimate a
stationary flux of the iron line at Earth from, e.g. the cloud Sgr
B2 which is at the level $\sim 10^{-4}$ ph cm$^{-2}$s$^{-1}$
\citep[see][]{dog_pasj1}, i.e just as observed. In terms of the
accretion model X-rays and subrelativistic protons are naturally
produced by star capture processes. Then we expect time variable
and stationary fluxes of 6.4 keV iron line emission from molecular
clouds generated by X-ray photons and hundred MeV protons
respectively.

{\it De-excitation gamma-ray lines.} Collisions of subrelativistic
nuclei with ambient matter can lead to nuclear excitation and
result in emission of de-excitation gamma-ray lines. These lines
may be a good tracer for subrelativistic cosmic rays, because the
line brightness can give us information about the amount of
subrelativistic particles. Assuming that the mean metallicity in
the GC region is two times higher than in the solar neighborhood,
 the total gamma-ray line flux below 8 MeV is
$1.1\times10^{-4}$~photons~cm$^{-2}$~s$^{-1}$. The most promising
lines for detection are those at 4.44 and $\sim$6.2~MeV, with a
predicted flux in each line of
$\approx$$10^{-5}$~photons~cm$^{-2}$~s$^{-1}$
\citep[see][]{dog_aa}. These lines should be broad, $\Delta
E_\gamma / E_\gamma$ of 3--4\%, which unfortunately renders their
detection with the {\it INTEGRAL} spectrometer unlikely but future
gamma-ray missions may be able to test these predictions.
Detection of the gamma-ray line emission produced by cosmic-ray
interactions in the interstellar medium would provide insightful
information on low-energy cosmic rays and give a significant
advance in the development of the theory of cosmic-ray origin.

\acknowledgements VAD and DOC are partly supported by the RFBR
grant 08-02-00170-a, the NSC-RFBR Joint Research Project ¹ RP09N04
and 09-02-92000-HHC-à. KSC is supported by a GRF grant under
HKU7014/07P.

\end{document}